\documentclass{raa07}

\usepackage{graphicx,times}             
\usepackage{natbib}
\usepackage{amssymb,amsmath}
\bibpunct{(}{)}{;}{a}{}{,}
\usepackage{CJK}
\usepackage[a4paper=true,pagebackref=true]{hyperref}
\hypersetup{colorlinks=true, linkcolor=green, anchorcolor=red, citecolor=blue, filecolor=red, pagecolor=red, urlcolor=red}
\PassOptionsToPackage{normalem}{ulem}

\begin{document}
\begin{CJK*}{UTF8}{gbsn}

  \title{A Two-limb Explanation for the Optical-to-infrared Transmission Spectrum of the Hot Jupiter HAT-P-32Ab}
   \volnopage{Vol.0 (20xx) No.0, 000--000}      
   \setcounter{page}{1}          
   \author{Xin-Kai Li (李馨凯) 
      \inst{1,2}
   \and Guo Chen (陈果)
      \inst{1,3}
   \and Hai-Bin Zhao (赵海斌)
      \inst{1,3}
   \and Hong-Chi Wang (王红池)
      \inst{4}
   }
   \institute{CAS Key Laboratory of Planetary Sciences, Purple Mountain Observatory, Chinese Academy of Sciences, Nanjing 210023, China; {\it guochen@pmo.ac.cn}\\
        \and
             School of Astronomy and Space Science, University of Science and Technology of China, Hefei 230026, China\\
        \and     
             CAS Center for Excellence in Comparative Planetology, Hefei 230026, China\\
        \and
             CAS Key Laboratory of Radio Astronomy, Purple Mountain Observatory, Chinese Academy of Sciences, Nanjing 210023, China\\
\vs\no
   {\small Received 2022 November 19; accepted 2022 December 21}}

\abstract{We present a new optical transmission spectrum of the hot Jupiter HAT-P-32Ab acquired with the Carnegie Observatories Spectrograph and Multiobject Imaging Camera (COSMIC) on the Palomar 200 inch Hale Telescope (P200). The P200/COSMIC transmission spectrum, covering a wavelength range of 3990--9390 {\AA}, is composed of 25 spectrophotometric bins with widths ranging from 200 to 400 {\AA} and consistent with previous transit measurements obtained in the common wavelength range. We derive a combined optical transmission spectrum based on measurements from five independent instruments, which, along with the 1.1--1.7 $\mu$m spectrum acquired by the Hubble Space Telescope and two Spitzer measurements, exhibits an enhanced scattering slope blueward of a relatively flat optical continuum, a water absorption feature at 1.4~$\mu$m, and a carbon dioxide feature at 4.4~$\mu$m. We perform Bayesian spectral retrieval analyses on the 0.3--5.1 $\mu$m transmission spectrum and find that it can be well explained by a two-limb approximation of $134^{+45}_{-33}\times$ solar metallicity, with a strongly hazy morning limb of $1134^{+232}_{-194}$~K and a haze-free evening limb of $1516^{+33}_{-44}$~K. This makes HAT-P-32Ab a promising target for James Webb Space Telescope to look for asymmetric signatures directly in the light curves.
}
   \authorrunning{Li et al.}            
   \titlerunning{Transmission spectrum of HAT-P-32Ab}  
   \maketitle
%
%


\section{Introduction}           
\label{sect:intro}
\end{CJK*}

The study of exoplanets is one of the fastest growing sub-disciplines in astronomy and planetary science. Observations and studies of exoplanet atmospheres have sprung up, and we have now discovered over 5200 exoplanets, among which over 3900 were discovered by the transit method (according to NASA Exoplanet Archive \footnote{\url{https://exoplanetarchive.ipac.caltech.edu/}}, as of 2022 November). During a transit, some of the stellar light will pass through the optically thin part of the planetary atmosphere, resulting in wavelength-dependent planetary radii with potential imprints of absorption and scattering features of planetary atmosphere at the terminator \citep{2000ApJ...537..916S}. It is feasible to retrieve the atmospheric properties from the observed transmission spectrum under certain model assumptions, e.g., line-by-line radiative transfer 1D model with parameterized temperature structure, chemical compositions, and clouds or hazes properties \citep{2009ApJ...707...24M}.

Close-in hot Jupiters are the most favorable targets for transmission spectroscopy with current instrumentation, which are gas giants with high temperatures, short orbital periods and extended atmospheres. The high atmospheric temperatures of hot Jupiters make them fantastic laboratories for unveiling the chemical abundances of giant planets. Through the analysis of the transmission spectrum of hot Jupiters, various species have been identified \citep{2019ARA&A..57..617M}, such as atomic metals including Na and K in the optical wavelengths due to their particularly prominent features at $\sim$589 and $\sim$768 nm \citep[e.g.,][]{2018Natur.557..526N,2018A&A...616A.145C}, and water vapor in the near infrared wavelengths for a water absorption feature centered at 1.4 $\mu$m \citep[e.g.,][]{2013ApJ...774...95D}. Among species found in the hottest hot Jupiters, gaseous TiO and VO may drive a temperature inversion in these planets \citep{2003ApJ...594.1011H}. The role of TiO/VO in these hottest atmospheres is still not clear, which could be depleted due to mechanisms such as deep-atmosphere or nightside cold trap, gravitational settling, photodissociation, thermal dissociation, and high C/O chemistry \citep{2009ApJ...699..564S,2009ApJ...699.1487S,2010ApJ...720.1569K,2012ApJ...758...36M,2013A&A...558A..91P,2018A&A...617A.110P}. The 3D global circulation models have been used to investigate the compositions, distributions, and formation of clouds and how they shape the transmission and emission spectra \citep{2016ApJ...828...22P,2019A&A...631A..79H,2021A&A...649A..44H}. Potential observational evidence for asymmetries resulting from atmospheric circulation has started to emerge through high-resolution Doppler spectroscopy \citep{2020Natur.580..597E,2022arXiv220313234V,2022A&A...668A..53C}.

One target of special interest is the highly inflated hot Jupiter HAT-P-32Ab, transiting a late-F-type star with a period of 2.15 days at a distance of 0.0343 AU discovered by \citet{2011ApJ...742...59H}. The planet has a mass of $0.585\pm0.031 M_\mathrm{Jup}$, a radius of $1.789\pm0.025 R_\mathrm{Jup}$, and an equilibrium temperature of $1801\pm18$ K, while the host star has a mass of $1.160\pm0.041 M_\odot$, a radius of $1.219\pm0.016 R_\odot$, and an effective temperature of $6269\pm64$ K \citep{2011ApJ...742...59H,2022A&A...657A...6C}. There is a resolved M1.5V companion, HAT-P-32B, at an angular separation of $2\farcs9$ \citep{2013AJ....146....9A}. HAT-P-32Ab is one of the best targets for transmission spectroscopy because of its large transit depth of more than 2\%, its large atmospheric scale height of about 1500 km, and a relatively bright host star ($V=11.4$~mag). 

Several observational studies have been conducted to reveal the atmospheric property of HAT-P-32Ab. Ground-based  binned spectrophotometry, from GMOS on Gemini North telescope (Gemini-N/GMOS) \citep{2013MNRAS.436.2974G}, MODS on Large Binocular Telescope (LBT/MODS) \citep{2016A&A...590A.100M}, and OSIRIS on Gran Telescopio Canarias (GTC/OSIRIS) \citep{2016A&A...594A..65N}, and multi-color broad-band photometry \citep{2016MNRAS.463..604M,2017AN....338..773M,2018MNRAS.474.5485T} all come to similar conclusions that HAT-P-32Ab has a flat featureless optical transmission spectrum, with a possible scattering slope at the blue-optical, indicative of a bimodal cloud distribution that consists of a Rayleigh-like haze and a gray cloud deck. The space observations, carried out with STIS and WFC3 on Hubble Space Telescope (HST), not only confirmed the enhanced Rayleigh scattering and the thick cloud deck \citep{2020AJ....160...51A}, but also revealed the presence of a water absorption feature at $\sim$1.4~$\mu$m \citep{2017AJ....154...39D,2020AJ....160...51A}. However, the full optical-to-infrared retrieval analysis performed by \citet{2020AJ....160...51A} cannot account for the shallower transit depths measured by Spitzer at 3.6 and 4.5 $\mu$m. In addition to transmission spectrum, dayside emission spectrum has also been acquired using ground-based $H$- and $K_S$-band photometry, Spitzer 3.6 and 4.5 $\mu$m photometry, and HST/WFC3 spectroscopy \citep{2014ApJ...796..115Z,2018MNRAS.474.1705N}, which shows no evidence of the $\sim$1.4~$\mu$m water feature but agrees with an isothermal atmosphere of $1995\pm 17$~K or an atmosphere with a modest thermal inversion.

In this work, we present a new optical transmission spectrum of HAT-P-32Ab observed by the Carnegie Observatories Spectrograph and Multiobject Imaging Camera \citep[COSMIC; ][]{1998PASP..110.1487K} on the Palomar 200 inch Hale Telescope (P200) and attempt to reconcile the existing discrepancy between data and model in the infrared. In Section \ref{sect:Obs}, we introduce the details of the observation and data reduction processes. In Section \ref{sect:lc}, we present the analyses on the white and spectroscopic light curves. In Section \ref{sect:retrieval}, we perform the Bayesian spectral retrieval analyses on HAT-P-32Ab's transmission spectrum. Finally, we discuss the implications of the retrieval results in Section \ref{sect:discussion} and draw conclusions in Section \ref{sect:conclusion}.

\section{Observations and Data Reduction}
\label{sect:Obs}
We obtained a transit time series of HAT-P-32Ab on the night of 2013 October 10 from 07:35 UT to 12:21 UT. The transit was observed with COSMIC installed at the prime focus of P200 located atop Palomar Mountain in north San Diego County, California. The spectroscopic mode of COSMIC has a field of view of $13\farcm65\times 13\farcm65$, equipped with a thinned, back-illuminated SITe $2048\times 2048$ CCD ($0\farcs4$ per pixel). A long-slit mask with a slit width of 12$''$ was created to simultaneously monitor the flux of HAT-P-32A ($V=11.3$~mag) and the reference star TYC 3281-957-1 ($V=11.0$~mag, $3\farcm2$ away). The 300 lines per mm grism was used to acquire the spectra, covering a wavelength range of 340--970 nm at a dispersion of $\sim$0.31 nm per pixel. An exposure time of 90 s was adopted, except for the first two taken with 60 and 120 s, resulting in a total of 85 frames. The long readout time of $\sim$117 s strongly reduced the duty cycle to 43\%. The mercury arc lamp was observed through a longslit mask with a slit width of 2$''$ for wavelength calibration. 

The raw spectral images were reduced following the methodology described in \citet{2021MNRAS.500.5420C} based on \texttt{IRAF} \citep{1993ASPC...52..173T} and customized IDL scripts, including corrections for overscan, bias, flat fields, sky background, and cosmic rays. The 1D spectra were extracted using the optimal extraction algorithm \citep{1986PASP...98..609H} with an aperture diameter of 21 pixels ($8\farcs4$), which minimized the scatter in the white-light curve. The white-light curve was created by summing the flux between 399 and 939 nm, while the spectroscopic light curves were created by binning the flux in a step of 20 nm, except for two 30 nm channels and one 40 nm channel at the longest wavelengths. The time stamp was extracted from the fits header and converted to Barycentric Julian Dates in Barycentric Dynamical Time \citep[$\mathrm{BJD_{TDB}}$; ][]{2010PASP..122..935E}.

\section{Light-curve Analysis}
\label{sect:lc}
\subsection{White-light curve}
\label{sect:wlc}

\begin{table}
\begin{center}
\caption[]{Parameters Estimated from the White-light Curve}\label{wlcpara}
 \begin{tabular}{lcc}
  \hline\noalign{\smallskip}
  \hline\noalign{\smallskip}
Parameter &  Prior      & Posterior Estimate                    \\
  \hline\noalign{\smallskip}
\textbf{$P$} [days]  & 2.15000815(fixed)         &  --       \\ 
$i$ [$^{\circ}$]  & $\mathcal{U}(80, 90)$            &   $88.57_{-0.60}^{+0.81}$                 \\
$a$/$R_*$         & $\mathcal{U}(3.0, 9.0)$          &   $6.202_{-0.069}^{+0.054}$                  \\
$R_p$/$R_*$       & $\mathcal{U}(0.10, 0.20)$        &   $0.1510_{-0.0011}^{+0.0011}$                  \\
$u_1$             & $\mathcal{N}(0.3520, 0.0340^2)$  &   $0.286_{-0.019}^{+0.021}$                  \\
$u_2$             & $\mathcal{N}(0.2991, 0.0182^2)$  &   $0.291_{-0.018}^{+0.018}$                  \\
$T_\mathrm{mid}$ $\mathrm{[MJD^{\color{blue}a}]}$    & $\mathcal{U}(76.8852, 76.9252)$  &   $76.90416_{-0.00011}^{+0.00011}$    \\
$\sigma_{\mathrm{w}}$ $[10^{-6}]$  & $\mathcal{U}(0.1, 5000)$      &   $287_{-27}^{+30}$             \\
$\ln A$             & $\mathcal{U}(-10, -1)$           &   $-5.0_{-1.1}^{+1.3}$   \\
$\ln\tau_1$        & $\mathcal{U}(-6, 5)$             &   $0.0_{-1.1}^{+1.2}$   \\
$\ln\tau_2$        & $\mathcal{U}(-5, 5)$             &   $2.7_{-1.1}^{+1.1}$    \\
$c_0$             & $\mathcal{N}(1.00587, 0.00039^2)$  &   $1.00587_{-0.00038}^{+0.00038}$    \\
$c_1$             & $\mathcal{N}(-0.0431, 0.0018^2)$   &   $-0.0431_{-0.0018}^{+0.0018}$   \\
$c_2$             & $\mathcal{N}(-0.400, 0.054^2)$   &   $-0.400_{-0.039}^{+0.040}$   \\
  \noalign{\smallskip}\hline
  \noalign{\smallskip}\hline
\end{tabular}
\end{center}
$^{\color{blue}a}\mathrm{MJD=BJD_{TDB}-2,456,500.}$
\end{table}
\begin{figure}
\centering
\includegraphics[width=\linewidth, angle=0]{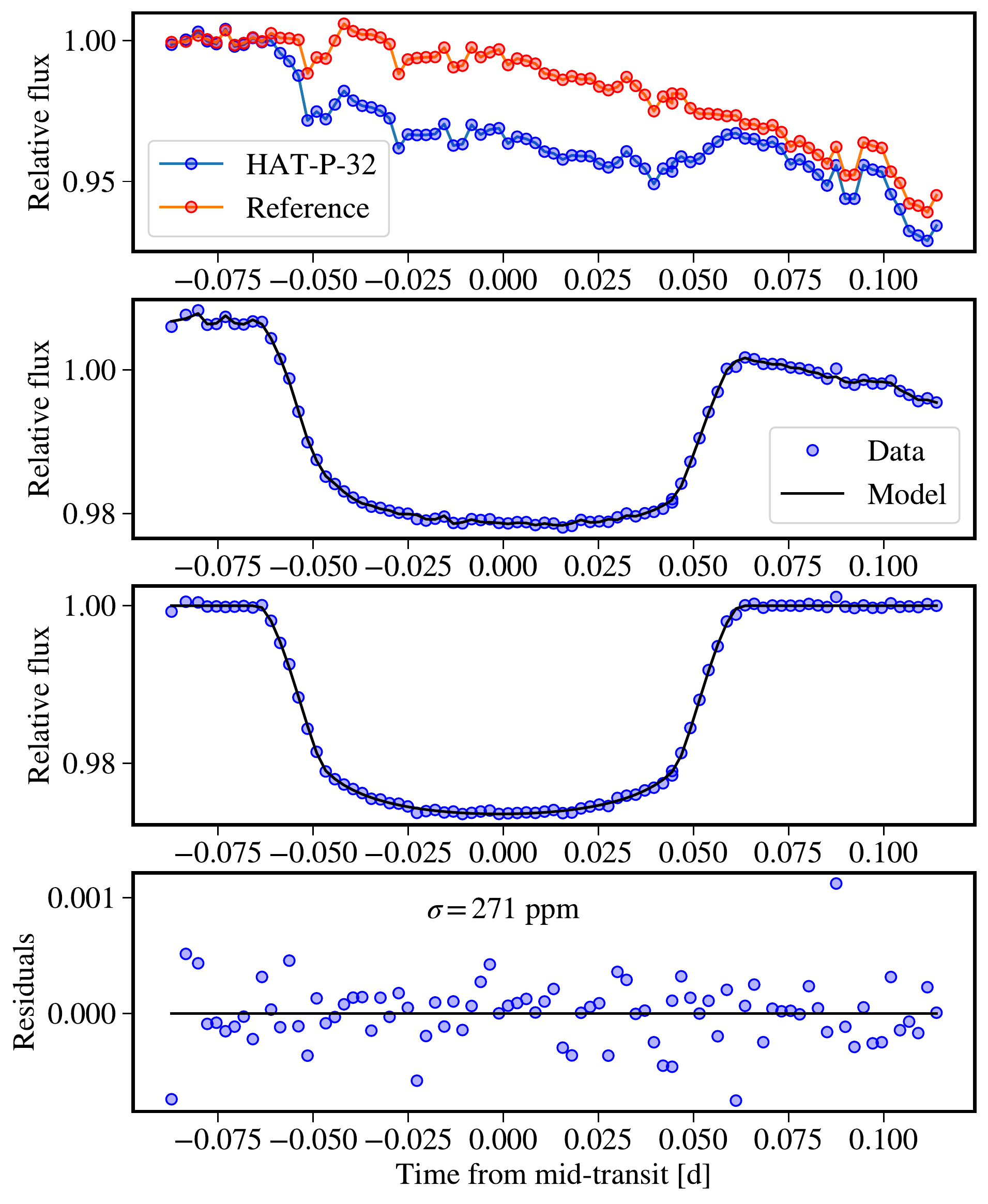}
\caption{White-light curve of HAT-P-32 observed by P200/COSMIC on the night of 2013 October 10. From top to bottom are (i) raw flux time series of HAT-P-32 and its reference star, (ii) raw white-light curve (i.e., normalized target-to-reference flux ratios), (iii) white-light curve corrected for systematics, and (iv) best-fit light-curve residuals. The best-fit models are shown in black.}
\label{whitelc}
\end{figure}
\begin{figure}
\centering
\includegraphics[width=\linewidth, angle=0]{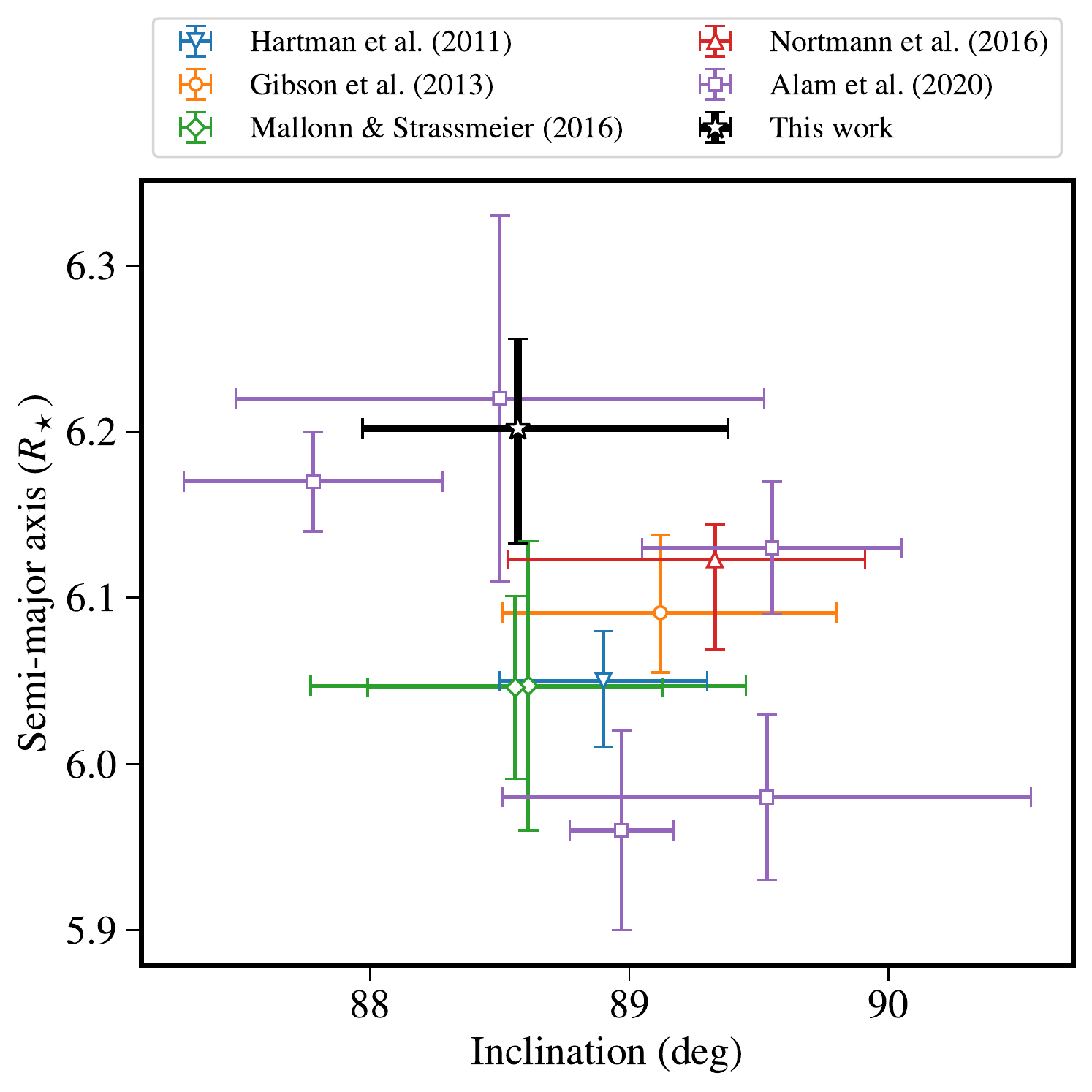}
\caption{Inclination and semi-major axis derived in this work compared to those in the literature.}
\label{inc_vs_ar}
\end{figure}

To model the transit light curve, we used the Python package \texttt{batman} \citep{2015PASP..127.1161K} configured with the quadratic limb-darkening law, which implements the analytic formalism from \citet{2002ApJ...580L.171M}. A circular orbit was assumed. The transit model $\mathcal{M}$ was parameterized by orbital period \citep[$P$; fixed to 2.15000815 days from][]{2021JBAA..131..359F}, orbital inclination ($i$), scaled semi-major axis ($a/R_\star$), radius ratio ($R_p/R_\star$), mid-transit time ($T_\mathrm{mid}$), and limb-darkening coefficients ($u_1$ and $u_2$). Since the close companion HAT-P-32B was not spatially resolved in our COSMIC observation, we revised the transit model as $\mathcal{M}'=(\mathcal{M}+f_d)/(1+f_d)$ to account for its dilution. The dilution flux ratio $f_d=F_B/F_A$ was calculated from the best-fit \texttt{PHEONIX} stellar template retrieved from the GTC/OSIRIS measurements presented by \citet{2016A&A...594A..65N} (see Table \ref{tstable_p200} in Appendix \ref{app1}).

To account for the correlated systematic noise in the observed light curve, we used the Python package \texttt{george} \citep{2015ITPAM..38..252A} to implement the Gaussian processes \citep[GPs; ][]{2006gpml.book.....R,2012MNRAS.419.2683G}. For the GP mean function, we adopted the transit model multiplied by a polynomial baseline, i.e., $\mathcal{M}'(c_0+c_1t+c_2t^2)$. For the GP covariance matrix, we used the product of two Mat\'ern $\nu=3/2$ kernels, with time ($t$) and spatial FWHM ($s$) as the input vectors, parameterized by an amplitude ($A$) and two characteristic length scales ($\tau_t$ and $\tau_s$). To account for potential underestimation of white noise, a jitter parameter $\sigma_j$ was added in the quadrature sum to the nominal flux uncertainties. 

To estimate the posterior distributions of the 13 free parameters ($i$, $a/R_\star$, $T_\mathrm{mid}$, $R_p/R_\star$, $u_1$, $u_2$, $c_0$, $c_1$, $c_2$, $A$, $\tau_t$, $\tau_s$, $\sigma_j$), we used the Python package \texttt{emcee} \citep{2013PASP..125..306F} to implement the affine invariant Markov Chain Monte Carlo (MCMC) ensemble sampler. In practice, the natural logarithmic values $\ln A$, $\ln\tau_t$, and $\ln\tau_s$ were used in the MCMC process. A total of 32 walkers were initialized and two short chains of 2000 steps were used for the ``burn-in'' phase. The final production was created after running a long chain of 50,000 steps that were thinned by every ten steps. We adopted uniform priors for all the parameters except for the baseline polynomial coefficients and the limb-darkening coefficients, which were controlled by normal priors. For the baseline polynomial coefficients, a second-order polynomial function was fitted to the out-of-transit flux and the resulting best-fit values and uncertainties were adopted as the normal priors. For the limb-darkening coefficients, the prior mean and sigma values were calculated from the ATLAS stellar models using the code developed by \citet{2015MNRAS.450.1879E}. 

The white light curve and best-fit model are shown in Figure~\ref{whitelc}. The best-fit light-curve residuals have a standard deviation of 270 ppm that is 3.6 times photon noise. The posteriors of free parameters are listed in Table \ref{wlcpara}. The derived transit parameters are in a broad agreement with those in the literature. Figure~\ref{inc_vs_ar} presents the comparison for $i$ and $a/R_\star$ between this work and the other transmission spectroscopy studies along with the discovery paper.

\subsection{Spectroscopic light curves}
\begin{figure}
\centering
\includegraphics[width=\linewidth, angle=0]{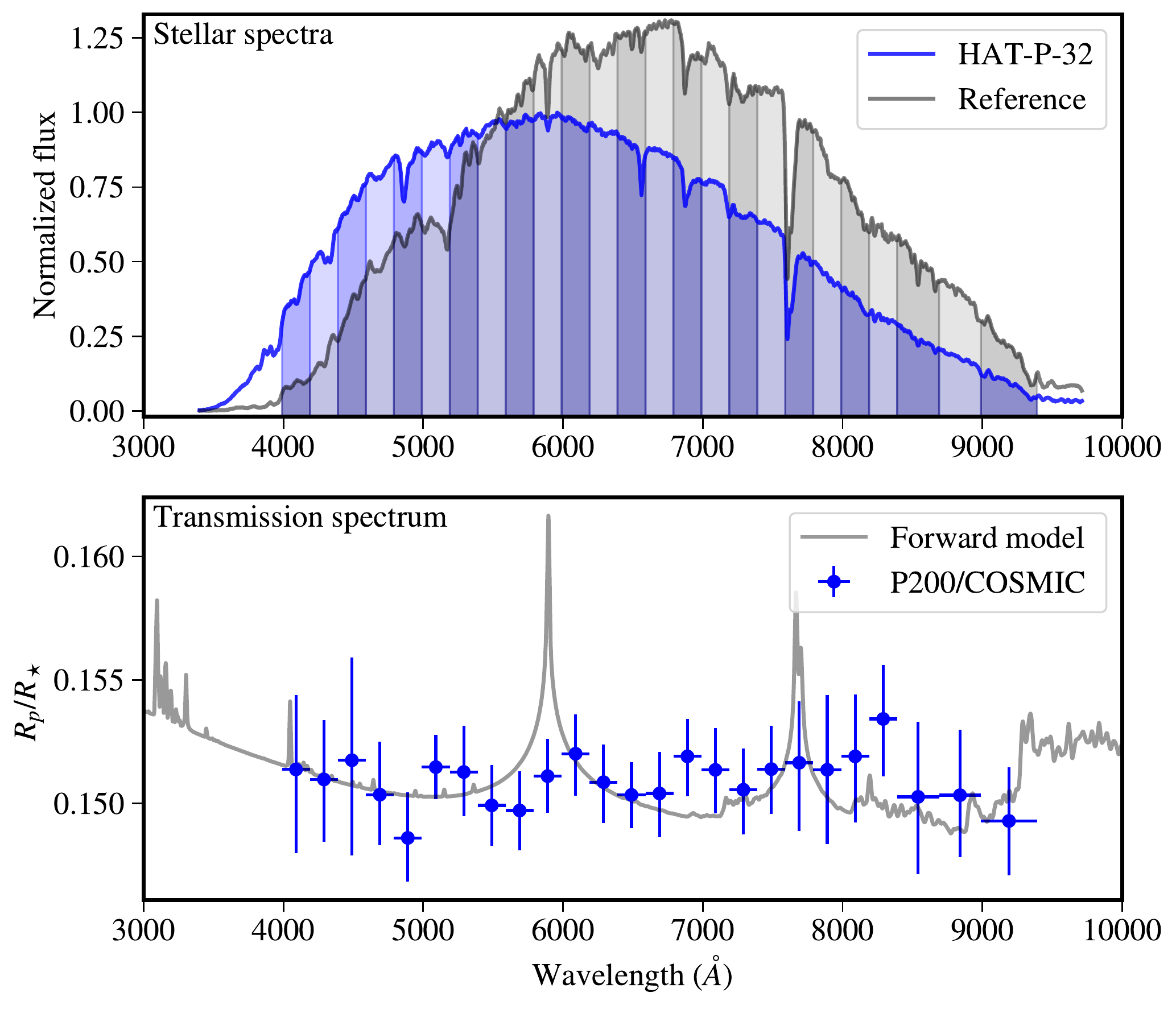}
\caption{The top panel shows the stellar spectra of HAT-P-32 and its reference star, along with passbands used in this work marked in shaded colors. The bottom panel presents the transmission spectrum of HAT-P-32Ab acquired with P200/COSMIC, compared to a 1800~K 1$\times$solar cloud-free fiducial model with Na and K but without TiO and VO. }
\label{st_pl_spec}
\end{figure}
\begin{figure*}
\centering
\includegraphics[width=\linewidth, angle=0]{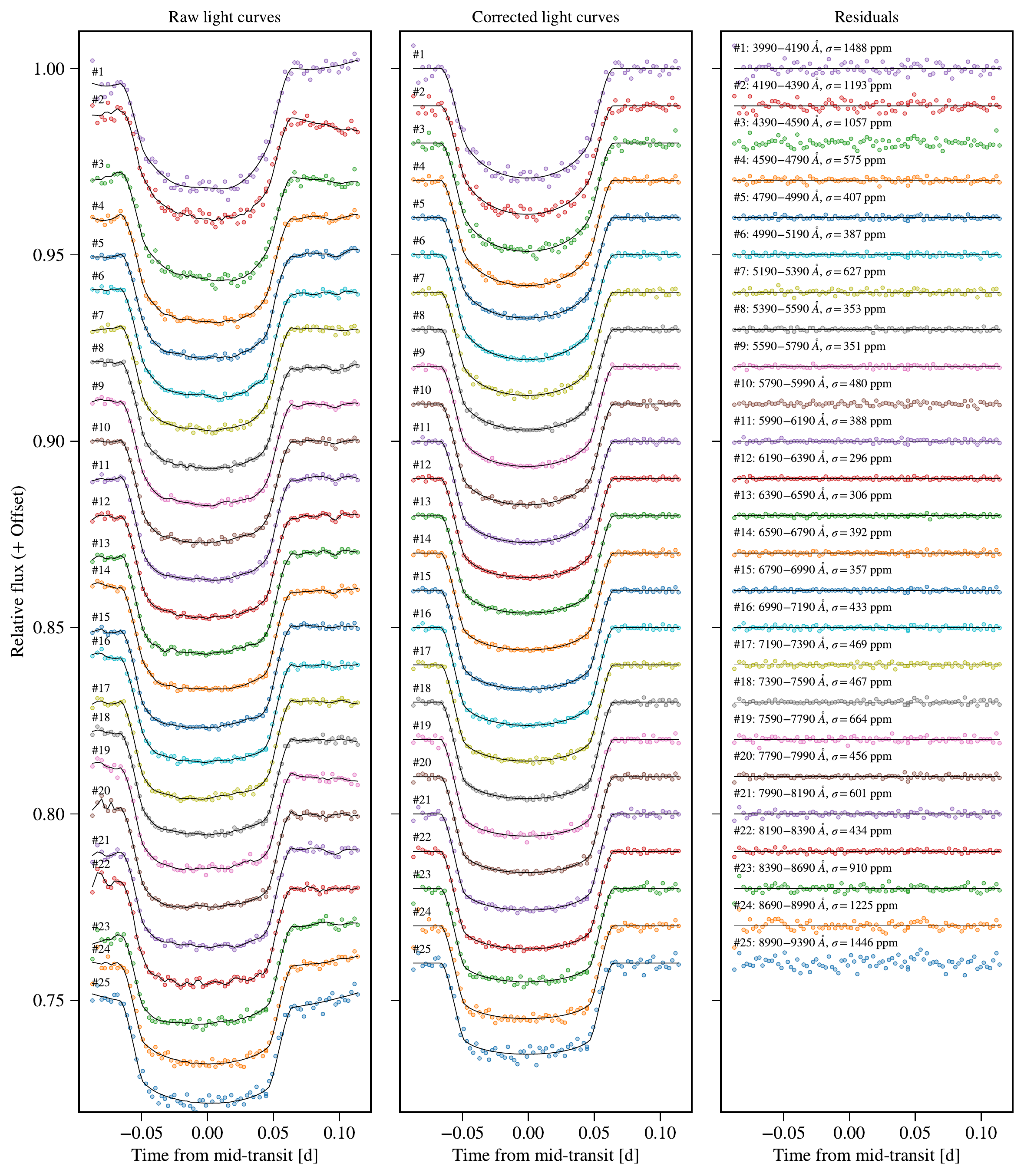}
\caption{Spectroscopic light curves of HAT-P-32 acquired with P200/COSMIC. From left to right are the original ones, the ones after removing the systematic models, and corresponding best-fit light-curve residuals. The black solid lines show the best-fit  models.}
\label{speclc}
\end{figure*}

Previous studies \citep{2018A&A...620A.142A,2020A&A...640A.134A} found that the shape of transmission spectrum could vary with the adopted orbital parameters due to the impact parameter degeneracy. In the case of HAT-P-32Ab, \citet{2020A&A...640A.134A} found negligible slope changes introduced by the impact parameter degeneracy. To compare with the results from HST, we adopted the nonlinear limb-darkening law, and fixed the four limb-darkening coefficients to the values interpolated from the Table 3 of \citet{2020AJ....160...51A}.

We modeled each individual spectroscopic light curve using the same method as that of the white-light curve, with the number of free parameters being reduced to eight ($R_p/R_\star$, $c_0$, $c_1$, $c_2$, $A$, $\tau_t$, $\tau_s$, $\sigma_j$) for each passband. We fixed $i$ and $a/R_\star$ to the values from \citet{2011ApJ...742...59H} as \citet{2020AJ....160...51A} did. We fixed $T_\mathrm{mid}$ to the value derived from the white-light curve (see Table \ref{wlcpara}). Since there was no significant common-mode noise, we did not apply the widely adopted common-mode removal technique to avoid potential underestimation of transit depth uncertainties \citep{2022A&A...664A..50J}. In the MCMC processes of the spectroscopic light curves, we also used 32 walkers and ran two short chains of 2000 steps for the ``burn-in'' phase, but created the final production after a long chain of 5000 steps without thinning. 

The adopted spectroscopic passbands are illustrated in the top panel of Figure~\ref{st_pl_spec}, while the derived wavelength dependent planet-to-star radius ratios, i.e., the transmission spectrum, are shown in the bottom panel and listed in Table \ref{tstable_p200}. The spectroscopic light curves and best-fit models are shown in Figure~\ref{speclc}. The standard deviations of the best-fit light-curve residuals for all the passbands are 0.8--2.4$\times$ photon noise, with a median value at 1.3$\times$.

\subsection{Transmission spectrum}
\label{transpec}
\begin{figure*}
\centering
\includegraphics[width=\linewidth, angle=0]{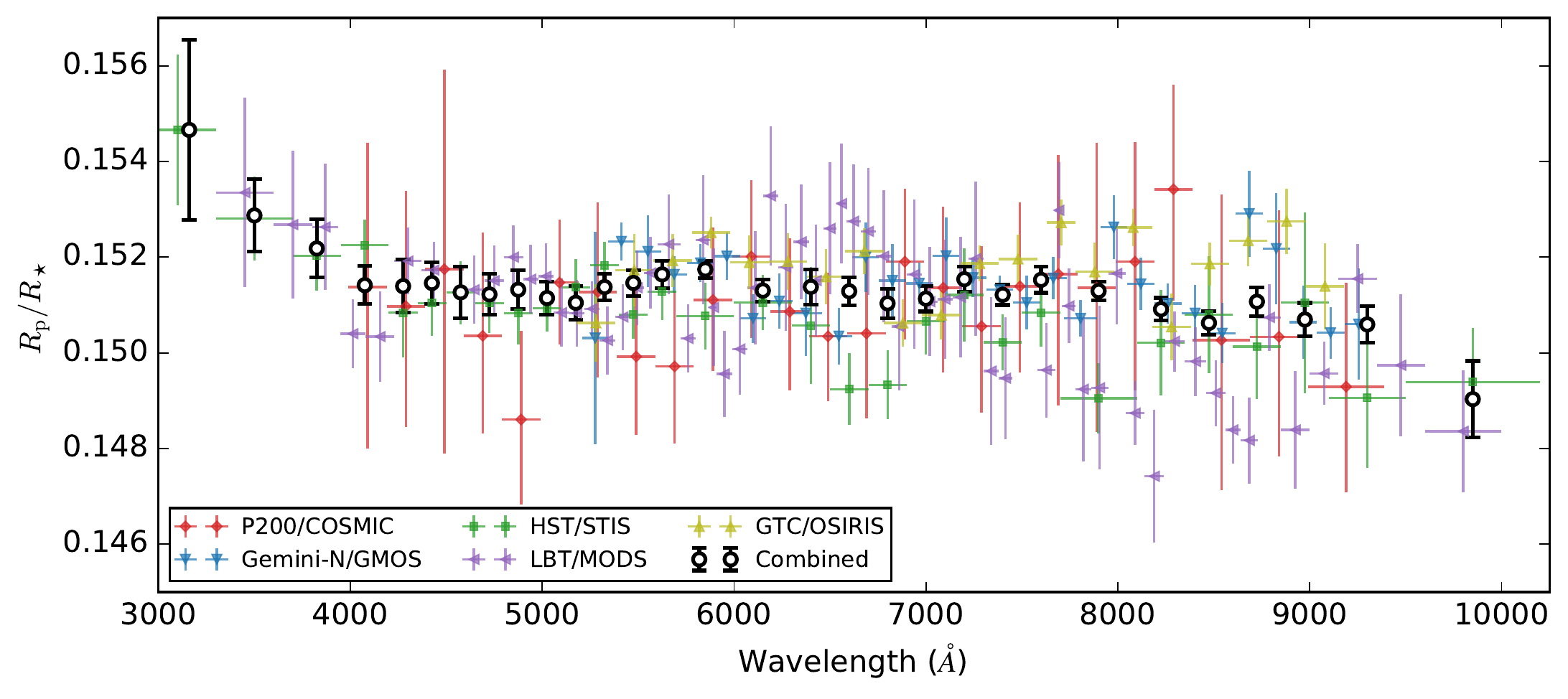}
\caption{Transmission spectra of HAT-P-32Ab acquired by Gemini-N/GMOS \citep[blue lower-triangles;][]{2013MNRAS.436.2974G}, LBT/MODS \citep[purple left-triangles;][]{2016A&A...590A.100M}, GTC/OSIRIS \citep[olive upper-triangles;][]{2016A&A...594A..65N}, HST/STIS \citep[green squares;][]{2020AJ....160...51A}, and P200/COSMIC (red diamonds). The combined optical transmission spectrum, i.e., weighted average of the five spectra resampled to the HST/STIS passbands, is shown in black circles.}
\label{comparison}
\end{figure*}

Our P200/COSMIC transmission spectrum is consistent with a flat line at a confidence level of 0.3$\sigma$ ($\chi^2=6.1$ for 24 degree of freedoms, hereafter dof). When compared to a sloped line in the form of $a_0+a_1\ln\lambda$, we obtain $\chi^2=6.0$ for 23 dof at a confidence level of 0.3$\sigma$. We further utilize the nested sampling algorithm \citep{2008MNRAS.384..449F} configured with 1000 live points to estimate the model evidence $\mathcal{Z}$ for model comparison. The Python package \texttt{PymultiNest} \citep{2014A&A...564A.125B} is used to implement the \texttt{MULTINEST} library \citep{2009MNRAS.398.1601F}. We interpret the model preference following the \citet{2008ConPh..49...71T} criteria, which categorizes the value of $\ln B_{10}=\ln\mathcal{Z}_1-\ln\mathcal{Z}_0$ into inconclusive ($|\ln B_{10}|<1$), weak ($1\leq |\ln B_{10}|<2.5$), moderate ($2.5\leq |\ln B_{10}|<5$), and strong ($|\ln B_{10}|\geq 5$). We find that the flat line has a model evidence higher than that of the sloped line ($\Delta\ln\mathcal{Z}=2.4$), indicating that it is not necessary to consider a non-flat model to explain the P200/COSMIC transmission spectrum.

Since the optical transmission spectrum of HAT-P-32Ab has also been acquired by Gemini-N/GMOS \citep{2013MNRAS.436.2974G}, LBT/MODS \citep{2016A&A...590A.100M}, GTC/OSIRIS \citep{2016A&A...594A..65N}, and HST/STIS \citep{2020AJ....160...51A}, we further investigate the consistency of transmission spectrum in the common wavelength range of different instruments. Figure \ref{comparison} shows the comparison between these five sets of optical transmission spectra. We downsample the other transmission spectra to the P200/COSMIC passbands with uncertainties propagated, and estimate a difference of $\chi^2=6.8$ (18 dof) for Gemini-N/GMOS, $\chi^2=13.7$ (24 dof) for LBT/MODS, $\chi^2=10.5$ (19 dof) for GTC/OSIRIS, $\chi^2=8.3$ (24 dof) for HST/STIS when compared to P200/DBSP. This reveals a good consistency between our spectrum and the others.

Given that different instruments might host specific systematics, we decide to combine the five spectra to alleviate the potential impact of specific systematics. We resample the other spectra onto the HST/STIS passbands and average these five spectra in the common passbands using the inverse square of uncertainties as weight. The combined optical transmission spectrum is presented in Table \ref{tstable_combined}. It features an enhanced slope, driven by HST/STIS and LBT/MODS, at $\lambda<4000$~{\AA}, and a relatively flat continuum at $4000<\lambda<9000$~{\AA} in the common wavelength range of all five spectra.

\section{Atmospheric retrieval}
\label{sect:retrieval}
In order to explore the atmospheric properties of HAT-P-32Ab, we perform Bayesian spectral retrieval analyses on the combined optical transmission spectrum derived in Sect. \ref{transpec} along with the HST/WFC3 and Spitzer measurements from \citet{2020AJ....160...51A}. We configure \texttt{PLATON} \citep{2019PASP..131c4501Z,2020ApJ...899...27Z} to implement forward modeling in the subsequent retrieval analyses, which conducts 1D radiative transfer to calculate the transmission spectrum of a hydrostatic atmosphere with equilibrium chemistry. 

The atmosphere is assumed to have an isothermal temperature of $T_\mathrm{iso}$, which is divided into 100 layers, log-equally spaced from 10$^3$ to 10$^{-9}$ bar, with the reference planet radius ($R_\mathrm{p,1bar}$) set at 1 bar. The gas absorption, collisional absorption, and scattering absorption are taken into account. The gas abundances are obtained from the equilibrium chemistry abundance grid pre-calculated by \texttt{GGChem} \citep{2018A&A...614A...1W}, which is a function of species name, temperature, pressure, metallicity ($Z$), and C/O ratio\footnote{To generate the input element abundances for \texttt{GGChem}, the abundances of elements above helium were scaled to $Z$ times their solar abundances and then the C abundance was varied to meet the requested C/O ratio (M. Zhang 2022, private communication).}. A total of 34 atomic and molecular species are considered, including: H, He, C, N, O, Na, K, $\mathrm{H}_2$, $\mathrm{H}_2$O, $\mathrm{CH}_4$, CO, $\mathrm{CO}_2$, $\mathrm{NH}_3$, $\mathrm{N}_2$, $\mathrm{O}_2$, $\mathrm{O}_3$, NO, $\mathrm{NO}_2$, $\mathrm{C}_2\mathrm{H}_2$, $\mathrm{C}_2\mathrm{H}_4$, $\mathrm{H}_2$CO, $\mathrm{H}_2$S, HCl, HCN, HF, MgH, OCS, OH, $\mathrm{PH}_3$, SiH, SiO, $\mathrm{SO}_2$, TiO, and VO. The gas opacities are calculated at a resolution of $\lambda/\Delta\lambda=10,000$, with line lists coming from ExoMol \citep{2018Atoms...6...26T}, HITRAN 2016 \citep{2017JQSRT.203....3G}, CDSD-4000 \citep{2011JQSRT.112.1403T}, \citet{2017ApJ...847..105R}, and NIST. The collisional absorption coefficients are taken from HITRAN \citep{2012JQSRT.113.1276R,2019Icar..328..160K}. The clouds are described as an optically thick cloud deck with a cloud-top pressure of $P_\mathrm{cloud}$. The hazes are parameterized to have a Rayleigh-like scattering with a slope of $\gamma$ at an amplitude of $A_\mathrm{scatt}$, i.e., $\sigma(\lambda)=A_\mathrm{scatt}\sigma_\mathrm{Rayleigh}\lambda^\gamma$. 

We adopt a planet mass of $0.585 M_\mathrm{J}$ \citep{2022A&A...657A...6C} in the forward model and use a stellar radius of $1.225 R_\odot$ \citep{2018MNRAS.474.5485T} to obtain the transit depth. We consider three model hypotheses in the following subsections and perform the retrieval analyses in the Bayesian framework. We configure \texttt{PymultiNest} with 250 live points to implement the nested sampling algorithm to estimate the model evidence and to explore the posterior distributions of free parameters. As summarized in Table \ref{tab_retrieval}, we adopt uniform or log-uniform priors for all the parameters. 
 
\begin{figure*}
\centering
\includegraphics[width=\linewidth, angle=0]{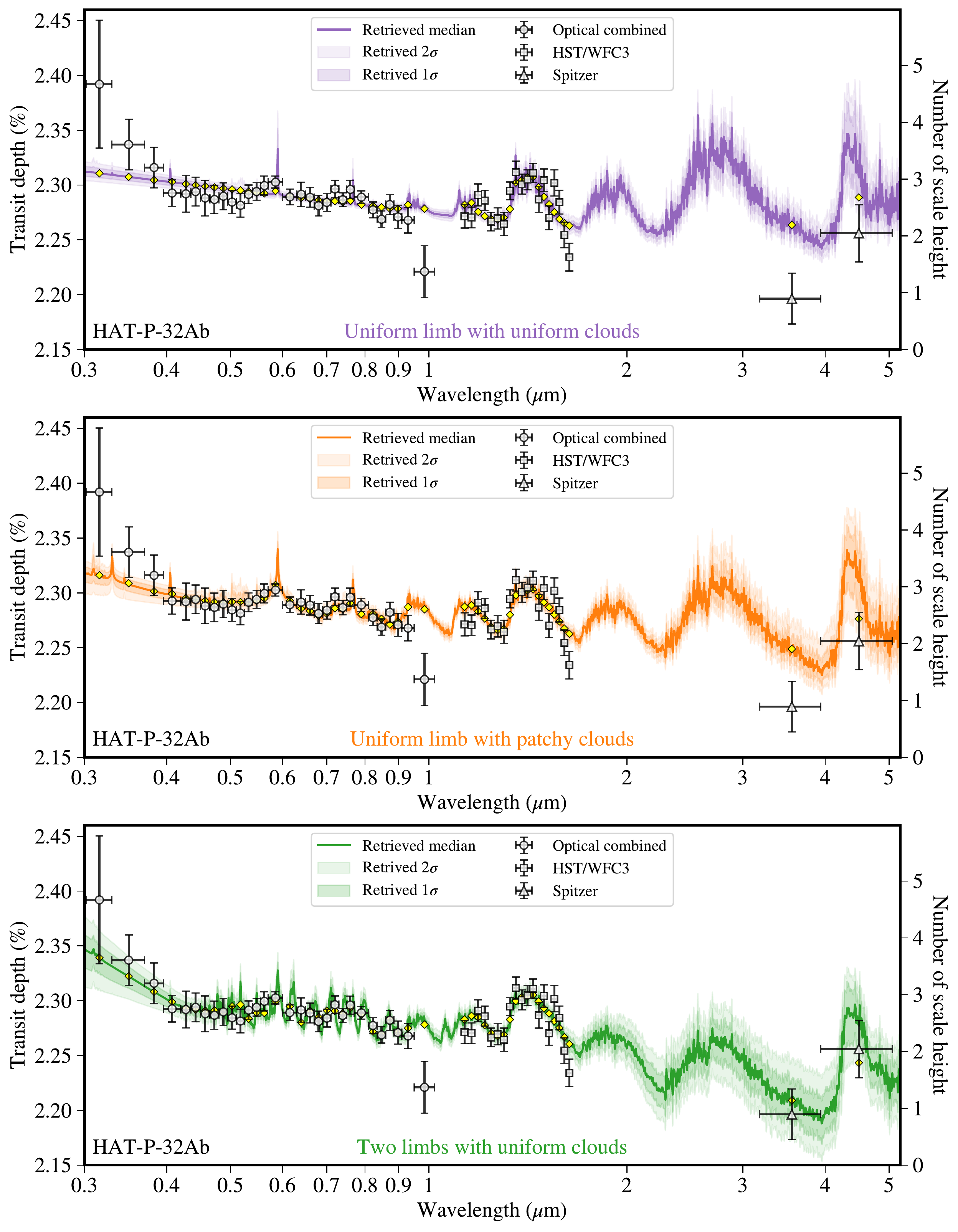}
\caption{Retrieved transmission spectra of HAT-P-32Ab from hypotheses of (i) uniform limb with uniform clouds (top panel), (ii) uniform limb with patchy clouds (middle panel), and (iii) two limbs with uniform clouds (bottom panel). The optical combined data are shown in circles, while the HST/WFC3 and Spitzer data from \citet{2020AJ....160...51A} are shown in squares and triangles, respectively. The median, 1$\sigma$, and 2$\sigma$ confidence intervals of retrieved models are shown in solid lines with shaded areas.}
\label{retrieved_ts}
\end{figure*}

\begin{table*}
\begin{center}
\caption[]{Parameter Estimation and Statistics from the Atmospheric Retrievals}\label{tab_retrieval}
 \begin{tabular}{llcccc}
  \hline\noalign{\smallskip}
  \hline\noalign{\smallskip}
Parameter              & Description                              &     Prior                & \multicolumn{3}{c}{Posterior} \\
                       &                                          &                          & Hypothesis 1            & Hypothesis 2            & Hypothesis 3 \\
  \hline\noalign{\smallskip}
$T_\mathrm{iso}$       & Atmospheric temperature (K)              & $\mathcal{U}(500, 2300)$ & $1130^{+139}_{-134}$    & $1288^{+121}_{-119}$     & --\\
$T_\mathrm{iso}^\mathrm{morn}$ & $T_\mathrm{iso}$ of morning limb & $\mathcal{U}(500, 2300)$ & --                      & --                      & $1134^{+232}_{-194}$\\
$T_\mathrm{iso}^\mathrm{even}$ & $T_\mathrm{iso}$ of evening limb & $\mathcal{U}(500, 2300)$ & --                      & --                      & $1516^{+33}_{-44}$\\
$\log P_\mathrm{cloud}$ & Log of cloud-top pressure (bar)         & $\mathcal{U}(-6, 2)$     & $-0.5^{+1.6}_{-1.7}$    & $-1.1^{+2.0}_{-2.3}$    & --\\
$\log P_\mathrm{cloud}^\mathrm{morn}$  & $\log P_\mathrm{cloud}$ of morning limb & $\mathcal{U}(-6, 2)$  & --          & --                      & $-1.8^{+2.5}_{-2.6}$\\
$\log P_\mathrm{cloud}^\mathrm{even}$  & $\log P_\mathrm{cloud}$ of evening limb & $\mathcal{U}(-6, 2)$  & --          & --                      & $-0.5^{+1.6}_{-1.6}$\\
$\gamma$               & Scattering slope                         & $\mathcal{U}(-20, 2)$    & $-1.1^{+0.2}_{-0.4}$    & $-3.0^{+0.8}_{-1.4}$    & --\\
$\gamma^\mathrm{morn}$ & $\gamma$ of morning limb                 & $\mathcal{U}(-20, 2)$    & --                      & --                      & $-17.1^{+3.4}_{-2.0}$\\
$\gamma^\mathrm{even}$ & $\gamma$ of evening limb                 & $\mathcal{U}(-20, 2)$    & --                      & --                      & $-15.7^{+4.5}_{-2.7}$\\
$\mathrm{log}A_\mathrm{scatt}$  & Log of scattering factor        & $\mathcal{U}(-4, 12)$    & $3.5^{+0.5}_{-0.5}$     & $6.3^{+0.6}_{-0.8}$     & --\\
$\mathrm{log}A_\mathrm{scatt}^\mathrm{morn}$ & $\mathrm{log}A_\mathrm{scatt}$ of morning limb & $\mathcal{U}(-4, 12)$  & --     & --             & $11.0^{+0.5}_{-0.7}$\\
$\mathrm{log}A_\mathrm{scatt}^\mathrm{even}$ & $\mathrm{log}A_\mathrm{scatt}$ of evening limb & $\mathcal{U}(-4, 12)$  & --     & --             & $-2.5^{+1.7}_{-1.0}$\\
$\mathrm{C/O}$         & Carbon-to-oxygen ratio                   & $\mathcal{U}(0.05, 2)$   & $0.30^{+0.21}_{-0.17}$  & $0.29^{+0.18}_{-0.15}$  & --\\
$\mathrm{C/O}^\mathrm{morn}$   & $\mathrm{C/O}$ of morning limb   & $\mathcal{U}(0.05, 2)$   & --                      & --                      & $0.82^{+0.76}_{-0.47}$\\
$\mathrm{C/O}^\mathrm{even}$   & $\mathrm{C/O}$ of evening limb   & $\mathcal{U}(0.05, 2)$   & --                      & --                      & $0.73^{+0.03}_{-0.04}$\\
$R_\mathrm{1bar}$      & Planet radius at 1 bar ($R_\mathrm{J}$)  & $\mathcal{U}(0.5, 2)$    & $1.704^{+0.016}_{-0.021}$ & $1.686^{+0.028}_{-0.028}$ & $1.704^{+0.014}_{-0.014}$\\
$\mathrm{log}Z$        & Log of metallicity ($Z_\odot$)           & $\mathcal{U}(-1, 3)$     & $0.96^{+0.78}_{-0.38}$  & $1.89^{+0.25}_{-0.28}$ & $2.13^{+0.13}_{-0.12}$\\
$\phi$                 & Fraction of cloud coverage               & $\mathcal{U}(0, 1)$      & --                      & $0.61^{+0.05}_{-0.07}$  & --\\
  \noalign{\smallskip}\hline
$\ln\mathcal{Z}$       & Model evidence                           & --                       & $366.1 \pm 0.3$         & $369.7 \pm 0.3$         & $374.2 \pm 0.3$\\
$\chi^2_\mathrm{MAP}$  & $\chi^2$ of maximum a posterior model    & --                       & 66.7                    & 58.3                    & 41.7\\
dof                    & Degree of freedom                        & --                       & 44                      & 43                      & 39\\
  \noalign{\smallskip}\hline
\end{tabular}
\end{center}
\end{table*}

\subsection{Hypothesis 1: uniform limb with uniform clouds}
By default, \texttt{PLATON} assumes an isothermal 1D atmosphere with uniform clouds (hereafter Hypothesis 1). The forward model consists of seven free parameters: $T_\mathrm{iso}$, $\log P_\mathrm{cloud}$, $\gamma$, $\log A_\mathrm{scatt}$, $\mathrm{C/O}$, $R_\mathrm{1bar}$, and $\log Z$. From the retrieval of Hypothesis 1, we obtain an isothermal temperature of $1130^{+139}_{-134}$~K, a subsolar C/O ratio of $0.30^{+0.21}_{-0.17}$, and a supersolar metallicity of $0.96^{+0.78}_{-0.38}$ dex that is slightly lower than the value retrieved from the HST and Spitzer measurements using the default setup of \texttt{PLATON} \citep{2020AJ....160...51A}. Our scattering slope of $-1.1^{+0.2}_{-0.4}$ is much shallower than the value of $-9.0^{+1.0}_{-0.6}$ reported by \citet{2020AJ....160...51A}, while our scattering amplitude of $10^{3.5\pm0.5}$ is much stronger than their $10^{1.0\pm0.4}$. This is mainly owing to the fact that the combined optical transmission spectrum is flatter than the HST/STIS spectrum within the wavelength range of 0.4--0.9~$\mu$m. However, we also note that the adopted priors and sampling algorithms might also introduce differences in the posterior estimates. While it is not clear what priors and sampling algorithms were used in \citet{2020AJ....160...51A}, the comparisons between parameters derived from our work and \citet{2020AJ....160...51A} should be taken with a grain of salt. We retrieve a loosely constrained cloud-top pressure with a 90\% lower limit of $\sim$2.6~mbar. The joint posterior distributions of all the free parameters are presented in Figure \ref{h1corner}.

The retrieved maximum a posterior (MAP) model can fit all the data at a 2.4$\sigma$ confidence level ($\chi^2_\mathrm{MAP}=66.7$ for 44 dof). As shown in the top panel of Figure \ref{retrieved_ts}, the retrieved models of Hypothesis 1 agree well with most of the data within 0.4--0.9 and 1.1--1.7~$\mu$m, but fail to fit a few data points, in particular those at the blue end and those in the mid infrared. The blue end shows an enhanced slope toward shorter wavelengths. The data point at 0.985~$\mu$m could be an outlier of the overall spectral shape given that the neighboring wavelength range of 0.5--0.9~$\mu$m is the average spectrum from five independent instruments while the one at 0.985~$\mu$m comes from the average of only two. In the mid infrared, the two Spitzer data points show clear offsets from the model predictions, and are also deviating from the overall optical-to-infrared trend. 

\subsection{Hypothesis 2: uniform limb with patchy clouds}
Three dimensional general circulation models (GCM) have predicted that the atmospheric circulation of hot Jupiters could induce asymmetries in temperature structure, chemistry, and clouds between the morning and evening limbs \citep[e.g.,][]{2009ApJ...699..564S,2016ApJ...828...22P,2019A&A...631A..79H}, resulting in different spectral signatures in the limb transmission spectra \citep{2010ApJ...709.1396F}. While it is possible to directly measure such asymmetries in the light curves if ultra-high photometric precision can be achieved \citep{2016A&A...589A..52V,2019ApJ...887..170P,2021AJ....162..165E}, we seek solutions through approximating the atmosphere by linear combinations of multi-sector 1D atmospheric models \citep{2016ApJ...820...78L,2017ApJ...845L..20K,2022ApJ...933...79W}.

In Hypothesis 2, we assume that the atmosphere is effectively composed of one clear sector and one cloudy sector, with $\phi$ being the faction of cloud coverage. Following \citet{2017MNRAS.469.1979M} and \citet{2021ApJ...913..114W}, the cloudy sector has a Rayleigh-like scattering haze above a gray cloud deck. This amounts to eight free parameters: $T_\mathrm{iso}$, $\log P_\mathrm{cloud}$, $\gamma$, $\log A_\mathrm{scatt}$, $\mathrm{C/O}$, $R_\mathrm{1bar}$, $\log Z$, and $\phi$. From the retrieval of Hypothesis 2, we obtain a cloud coverage of $61^{+5}_{-7}$~\%, an isothermal temperature of $1288^{+121}_{-119}$~K, a subsolar C/O ratio of $0.29^{+0.18}_{-0.15}$, and a supersolar metallicity of $1.89^{+0.25}_{-0.28}$ dex. For the haze property, the scattering slope of $-3.0^{+0.8}_{-1.4}$ is slightly steeper, while the scattering amplitude increases significantly to $10^{6.3^{+0.6}_{-0.8}}$. The cloud-top pressure is again loosely constrained, with a 90\% lower limit of $\sim$0.1~mbar. The joint posterior distributions of all the free parameters are presented in Figure \ref{h2corner}.

The retrieved models of Hypothesis 2 are shown in the middle panel of Figure \ref{retrieved_ts}. The MAP model of Hypothesis 2 can fit all the data at the 1.9$\sigma$ confidence level ($\chi^2_\mathrm{MAP}=58.3$ for 43 dof), slightly better than Hypothesis 1. The major improvement in Hypothesis 2 occurs in the optical, where the new retrieval suggests the presence of pressure-broadened line wings of Na and K. The strong scattering amplitude introduced by the haze in the cloudy sector also slightly moves downwards the model in the mid-infrared. However, the enhanced slope at the blue end and the Spitzer data points are still not well explained.

\begin{figure*}
\centering
\includegraphics[width=\linewidth, angle=0]{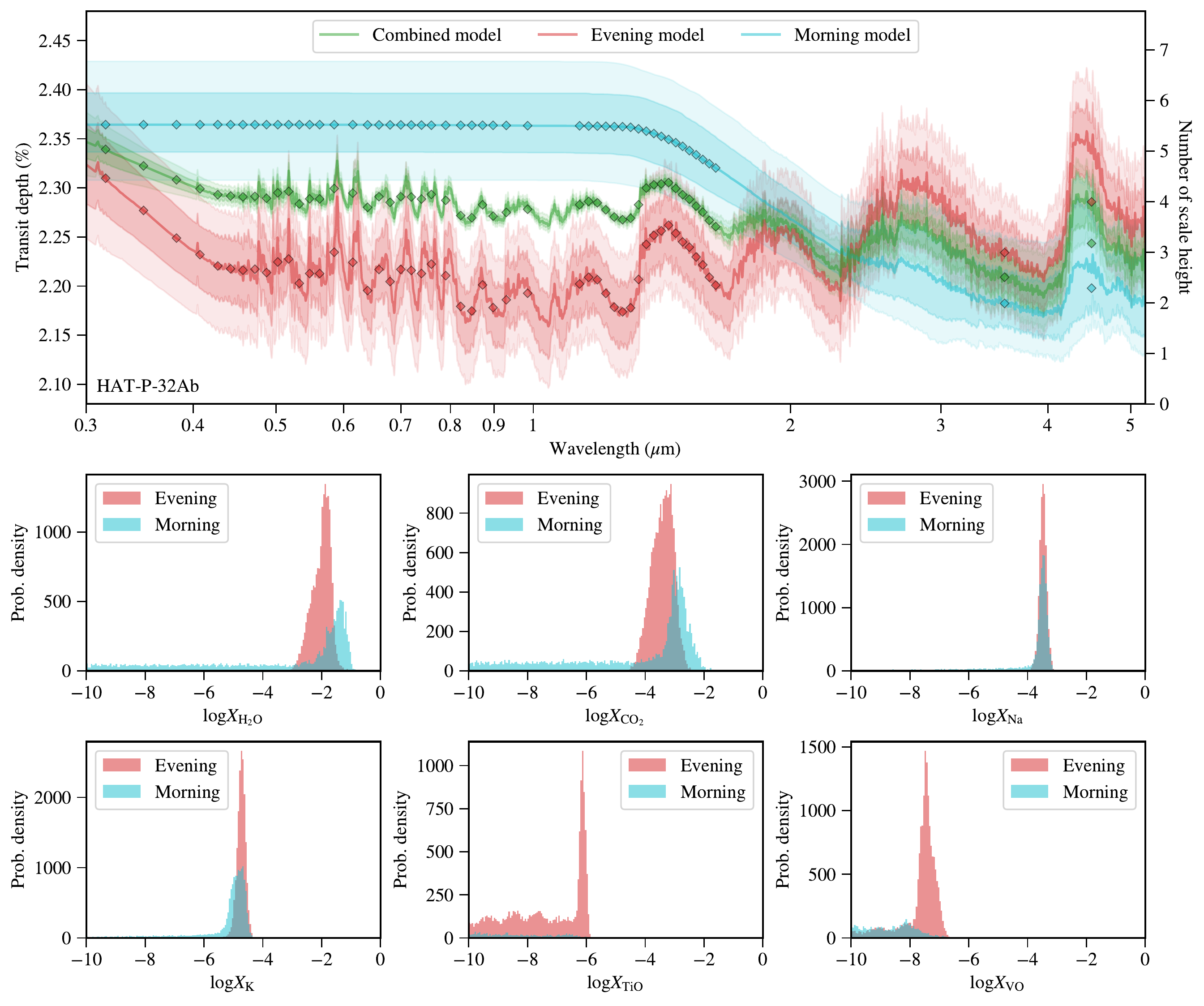}
\caption{The first row presents the confidence regions of the derived morning-limb and evening-limb transmission spectra based on the two-limb retrieval in Hypothesis 3. The second and third rows show the marginalized posterior distributions of the volume mixing ratios of the major spectral tracing species retrieved in Hypothesis 3, including H$_2$O, CO$_2$, Na, K, TiO, and VO.}
\label{species_posteriors}
\end{figure*}

\subsection{Hypothesis 3: two limbs with uniform clouds}
While the patchy clouds could result in asymmetries in the limb, as assumed in Hypothesis 2, it is also possible that the morning limb and evening limb are indeed asymmetric. Similar to the retrieval frameworks adopted in \citet{2021AJ....162..165E} and \citet{2022ApJ...933...79W}, we assume in Hypothesis 3 that the atmosphere can be equivalent to a cooler morning limb and a warmer evening limb with equal weights, which have separate isothermal temperatures ($T_\mathrm{iso}^\mathrm{morn}$ and $T_\mathrm{iso}^\mathrm{even}$). For both limbs, uniform clouds are adopted individually, each composed of a Rayleigh-like scattering haze above a gray cloud deck. The 3D GCM studies suggest that planets with intermediate temperatures (1400--1800~K) could have homogeneous mean molecular weight but intermittent C/O ratio across observable planet disk \citep{2022arXiv220805562H}. Therefore, the C/O ratio is considered to be limb-dependent, while the metallicity is assumed to be the same for both limbs. Consequently, there are 12 free parameters: $T_\mathrm{iso}^\mathrm{morn}$, $T_\mathrm{iso}^\mathrm{even}$, $\log P_\mathrm{cloud}^\mathrm{morn}$, $\log P_\mathrm{cloud}^\mathrm{even}$, $\gamma^\mathrm{morn}$, $\gamma^\mathrm{even}$, $\log A_\mathrm{scatt}^\mathrm{morn}$, $\log A_\mathrm{scatt}^\mathrm{even}$, $\mathrm{C/O}^\mathrm{morn}$, $\mathrm{C/O}^\mathrm{even}$, $R_\mathrm{1bar}$, and $\log Z$.

From the retrieval of Hypothesis 3, we obtain a supersolar metallicity of $2.13^{+0.13}_{-0.12}$ dex. The retrieved C/O ratios are supersolar and almost the same for both limbs ($\sim$0.7), although the constraints are tighter for the evening limb while looser for the morning limb. The retrieved temperatures are $1134^{+232}_{-194}$~K for the morning limb and $1516^{+33}_{-44}$~K for the evening limb, resulting in an evening-morning difference of $\Delta T=373^{+192}_{-224}$~K. This is consistent with the evening-averaged and morning-averaged temperature profiles within 1--10~mbar derived from cloud-free GCM simulations for another hot Jupiter with similar physical properties \citep[WASP-17b;][]{2016ApJ...821....9K}. The scattering slopes are loosely constrained, with a 90\% upper limit of $\gamma<-12.5$ for the morning limb and $\gamma<-9.8$ for the evening limb. The scattering amplitudes are much stronger in the morning limb ($A_\mathrm{scatt}=10^{11.0^{+0.5}_{-0.7}}$) than in the evening limb ($A_\mathrm{scatt}=10^{-2.5^{+1.7}_{-1.0}}$), indicating that the morning limb is dominated by a very strong haze and that the evening limb is almost haze free. Like Hypotheses 1 and 2, the cloud-top pressure is also loosely constrained in Hypothesis 3, with a 90\% lower limit of 0.02~mbar in the morning limb and 4~mbar in the evening limb. However, the lower limit does indicate that the cloud deck in the evening might be deeper. The joint posterior distributions of all the free parameters are presented in Figure \ref{h3corner}.

The retrieved models of Hypothesis 3 are shown in the bottom panel of Figure \ref{retrieved_ts}. The MAP model of Hypothesis 3 can fit all the data at the 0.9$\sigma$ confidence level ($\chi^2_\mathrm{MAP}=41.7$ for 39 dof), better than both Hypotheses 1 and 2. The enhanced slope at the blue end and the Spitzer data points can be reasonably explained by the models of Hypothesis 3. The optical spectral features in the MAP model are dominated by the opacities of TiO, VO, MgH, Na, and K in the evening limb. If the data point at 0.985~$\mu$m is excluded as an outlier, the values of $\chi^2_\mathrm{MAP}$ are reduced to 61.3, 51.1, and 35.8 for Hypotheses 1, 2, and 3, respectively, corresponding to goodness of fit at 2.0$\sigma$, 1.3$\sigma$, and 0.5$\sigma$.

\section{Discussion}
\label{sect:discussion}

Our Bayesian spectral retrieval analyses reveal that the current 0.3--5.1~$\mu$m transmission spectrum data set of HAT-P-32Ab strongly favors the hypothesis of two limbs with uniform clouds as opposed to the hypotheses of uniform limb with either uniform clouds or patchy clouds. The retrieved models from the two-limb hypothesis can fit the current data set reasonably well, indicating that the atmosphere of HAT-P-32Ab can be equivalent to a warmer hazy-free component and a cooler hazy component, which we attribute to the evening and morning limbs. Although the retrieved C/O ratios are almost the same ($\sim$0.7) on both limbs, it is largely unconstrained in the morning limb. Figure \ref{species_posteriors} presents the confidence regions of the derived morning-limb and evening-limb transmission spectra based on the two-limb retrieval. It also shows the marginalized limb-dependent posteriors of the major spectral tracers that are derived from our equilibrium chemistry retrieval, of which Na, K, TiO, and VO contribute to the optical wavelengths, H$_2$O contributes to the infrared wavelengths, and CO$_2$ contributes to the Spitzer bands. The most evident abundance changes occur in TiO and VO, which are likely depleted in the cooler morning limb due to either condensation or nightside cold trap, consistent with GCM predictions \citep{2016ApJ...828...22P,2022arXiv220805562H}. 

Based on the assumption of equilibrium chemistry, the atmospheric metallicity of HAT-P-32Ab is constrained to be $134^{+45}_{-33}$ times solar metallicity, which is strongly enhanced over its solar-metallicity host star \citep[$\mathrm{[Fe/H]}=-0.04\pm 0.08$;][]{2011ApJ...742...59H} and much stronger than the observed planet mass-metallicity enrichment trend at the mass of 0.585~$M_\mathrm{J}$ \citep[e.g.,][]{2014ApJ...793L..27K,2017Sci...356..628W,2019ApJ...887L..20W}. The enrichment of atmospheric metallicity toward lower planet mass has been suggested as a potential link to the core-accretion planet formation theory \citep{2011ApJ...736L..29M,2013ApJ...775...80F,2014A&A...566A.141M,2016ApJ...831...64T}. However, \citet{2019ApJ...887L..20W} found that the enrichment trends could differ if it was based on different species (e.g., CH$_4$, H$_2$O), suggesting that the equilibrium chemistry assumption might not work in general. Future observations with James Webb Space Telescope (JWST) that cover a variety of molecular species in the infrared wavelengths will enable us to answer whether or not equilibrium chemistry is reasonable for HAT-P-32Ab. 

The powerful capability of JWST could also enable a direct measurement of the transmission spectrum for each individual limb, which can independently confirm whether the morning-evening asymmetries are present. The morning-to-evening transit depth differences derived from our retrieval have a maximum value of $\sim$990~ppm within 0.3--5.1~$\mu$m, which could induce an asymmetric signature as large as $\sim$400~ppm during ingress or egress in the light-curve residuals when compared to the symmetric light-curve model depending on the orientation of the semi-circles of the evening and morning limbs. Asymmetric signatures of such amplitude are easily detectable by JWST according to the simulations on the hot Jupiter HAT-P-41b orbiting a star of similar spectral type and brightness \citep{2021AJ....162..165E}. 

We note that the major driver to favor the two-limb hypothesis in this work could probably come from the enhanced slope at the blue end and the two Spitzer data points that have a potential CO$_2$ spectral signature but are at a lower level than the 0.3--1.7~$\mu$m wavelength range. The offset between the two Spitzer data points and other wavelengths could also come from contamination of stellar activity or instrumental biases. The five-season photometric monitoring reveals that HAT-P-32A is constant on night-to-night timescales within the precision of $\sim$2~mmag and likely to be constant on year-to-year timescales \citep{2018MNRAS.474.1705N,2020AJ....160...51A}. The consistency among the optical transmission spectra (0.5--0.9~$\mu$m) independently acquired by five instruments and the consistency between optical (0.3--0.9~$\mu$m) and near-infrared (1.1--1.7~$\mu$m) both confirm the inactive nature of HAT-P-32A. The large instantaneous wavelength coverage of JWST will further confirm whether the downward mid-infrared spectral signature of CO$_2$ is of instrumental origin or a sign of morning-evening asymmetry \citep{2022arXiv220811692T}.

\section{Conclusions}
\label{sect:conclusion}
We obtained an optical transmission spectrum for the hot Jupiter HAT-P-32Ab within the wavelength range of 399--939~nm using P200/COSMIC. We derived a combined optical transmission spectrum by weighted averaging the measurements from five independent instruments including HST/STIS, Gemini-N/GMOS, LBT/MODS, GTC/OSIRIS, and P200/COSMIC. We performed Bayesian spectral retrievals on the combined optical spectrum along with the HST/WFC3 and Spitzer measurements, with the hypotheses of (i) uniform limb with uniform clouds, (ii) uniform limb with patchy clouds, and (iii) two limbs with uniform clouds. We conclude that: 
\begin{itemize}
\item[1.] The current 0.3--5.1~$\mu$m transmission spectrum of HAT-P-32Ab is characterized by an enhanced scattering slope at the blue-optical, a relatively flat continuum but consistent with spectral signatures of TiO, VO, Na, K, and MgH in the optical band, a water absorption feature at 1.4~$\mu$m, and a CO$_2$ absorption feature at 4.4~$\mu$m.
\item[2.] The current data set of HAT-P-32Ab reveals an atmosphere of high metallicity ($\mathrm{[Fe/H]}=2.13^{+0.13}_{-0.12}$), and can be well explained by a two-limb approximation, with the warmer evening limb being haze-free and the cooler morning limb being strongly hazy. The morning-evening temperature difference of $373^{+192}_{-224}$~K is consistent with the GCM predictions.
\item[3.] HAT-P-32Ab is a prior target for follow-up observations with JWST transmission spectroscopy. The inferred morning and evening limbs, if confirmed, will enable direct measurements of limb spectra through asymmetric light-curve modeling.
\end{itemize}

\begin{acknowledgements}
G.C. acknowledges the support by the B-type Strategic Priority Program of the Chinese Academy of Sciences (grant No.\,XDB41000000), the National Natural Science Foundation of China (grant Nos. 42075122, 12122308), Youth Innovation Promotion Association CAS (2021315). H.Z. thanks the Space debris and NEO research project (grant Nos. KJSP2020020204, KJSP2020020102), Civil Aerospace pre-research project (grant No. D020304). G.C. and H.Z. also thank the Minor Planet Foundation. 
The authors would like to thank the anonymous referee for the constructive comments on the manuscript, and Carolyn Heffner, Kajsa Peffer, Kevin Rykoski, and Jennifer Milburn for their great supports during the observations. 
This research uses data obtained through the Telescope Access Program (TAP), which has been funded by the TAP member institutes. 
Observations obtained with the Hale Telescope at Palomar Observatory were obtained as part of an agreement between the National Astronomical Observatories, Chinese Academy of Sciences, and the California Institute of Technology.
\end{acknowledgements}

\bibliographystyle{raa}
\bibliography{ref}

\appendix 

\section{Additional Tables and Figures}
\label{app1}
Table~\ref{tstable_p200} presents the transmission spectrum of HAT-P-32Ab measured by P200/COSMIC. Table~\ref{tstable_combined} presents the combined optical transmission spectrum of HAT-P-32Ab, which is a weighted average of transmission spectra measured by five independent instruments, including Gemini-N/GMOS, LBT/MODS, GTC/OSIRIS, HST/STIS, and P200/COSMIC.

\begin{figure*}
\centering
\includegraphics[width=\linewidth, angle=0]{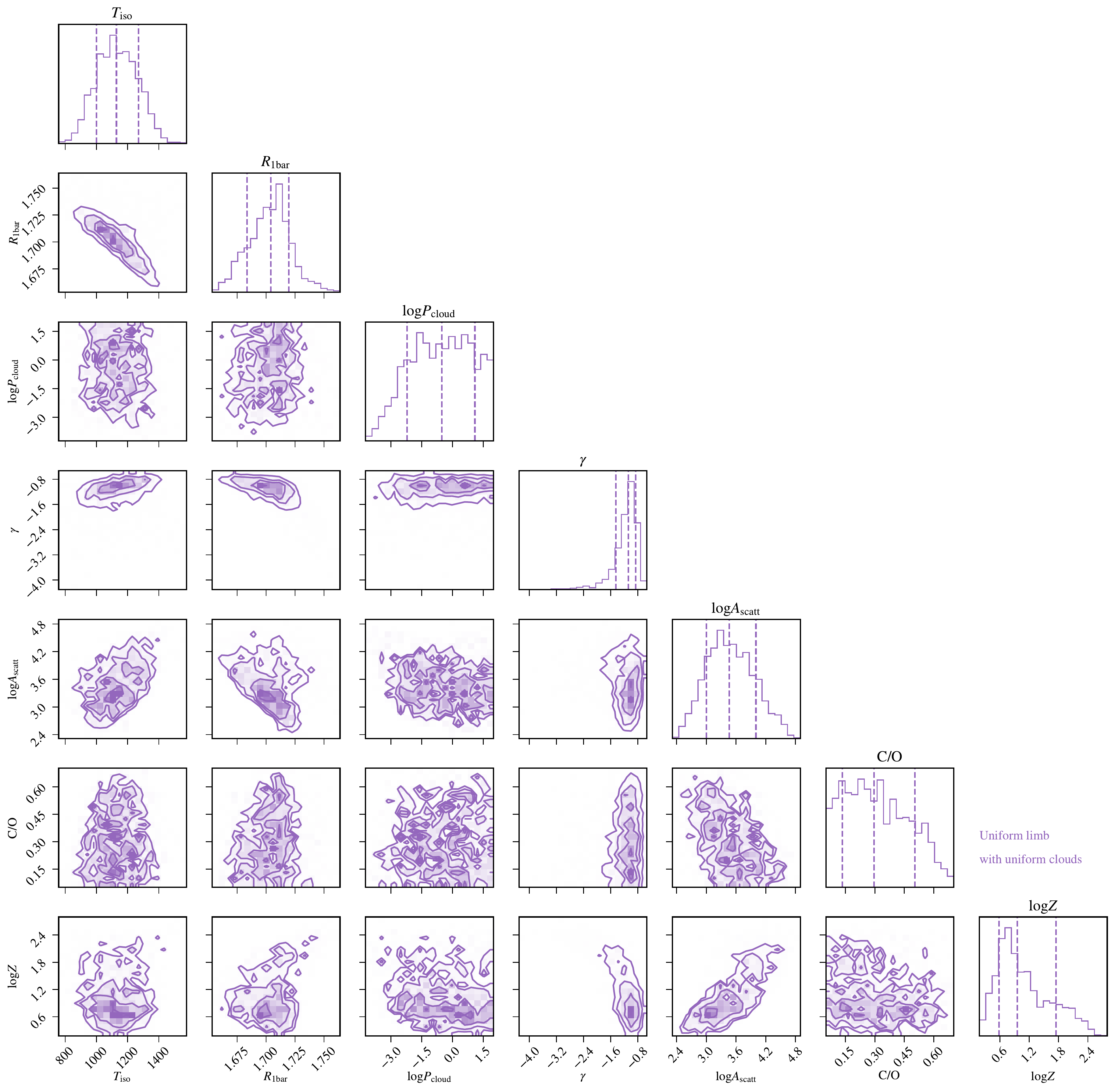}
\caption{Corner plots of the retrieval of Hypothesis 1, where uniform limb with uniform clouds is assumed.}
\label{h1corner}
\end{figure*}

\begin{figure*}
\centering
\includegraphics[width=\linewidth, angle=0]{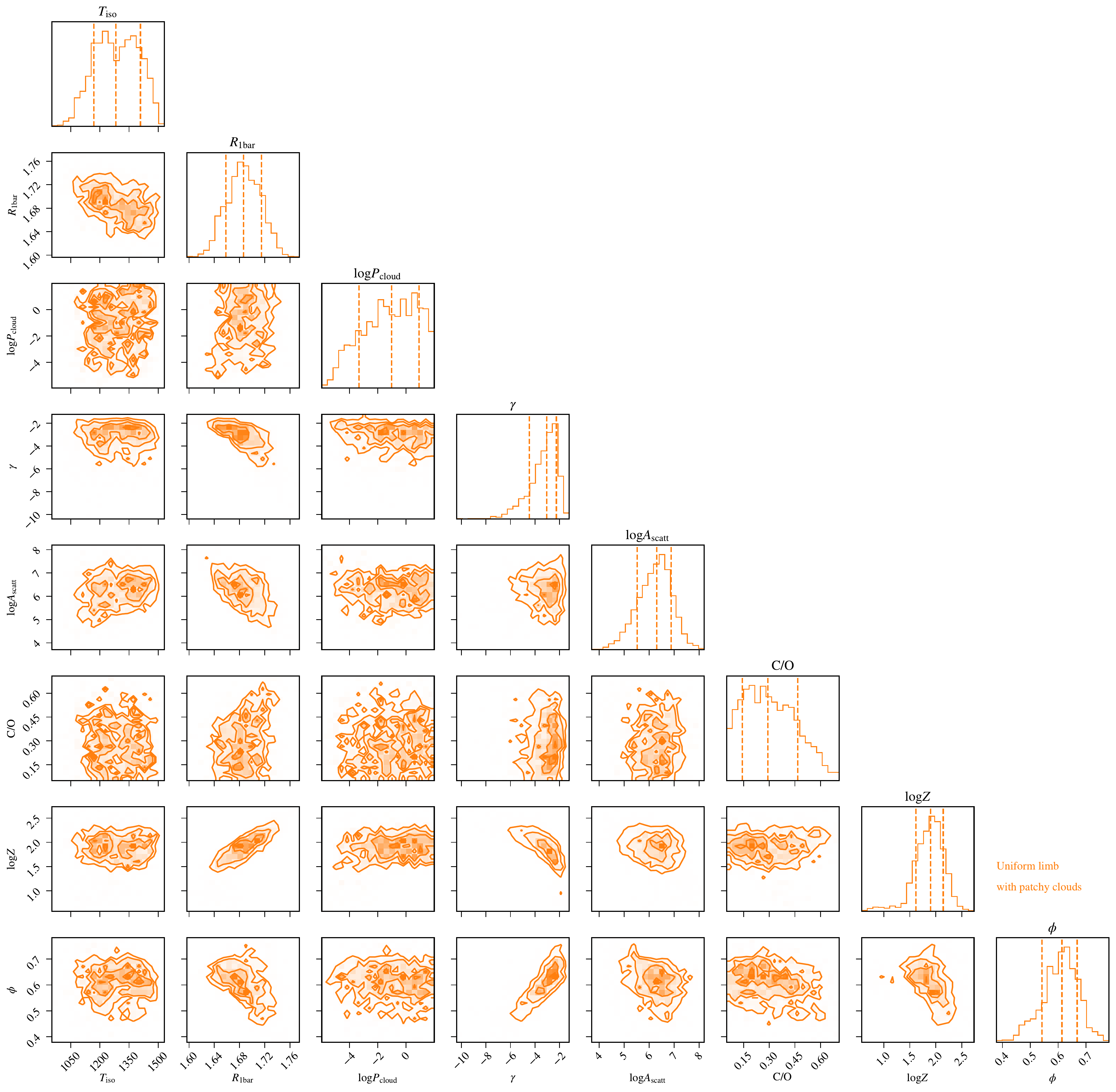}
\caption{Corner plots of the retrieval of Hypothesis 2, where uniform limb with patchy clouds is assumed.}
\label{h2corner}
\end{figure*}

\begin{figure*}
\centering
\includegraphics[width=\linewidth, angle=0]{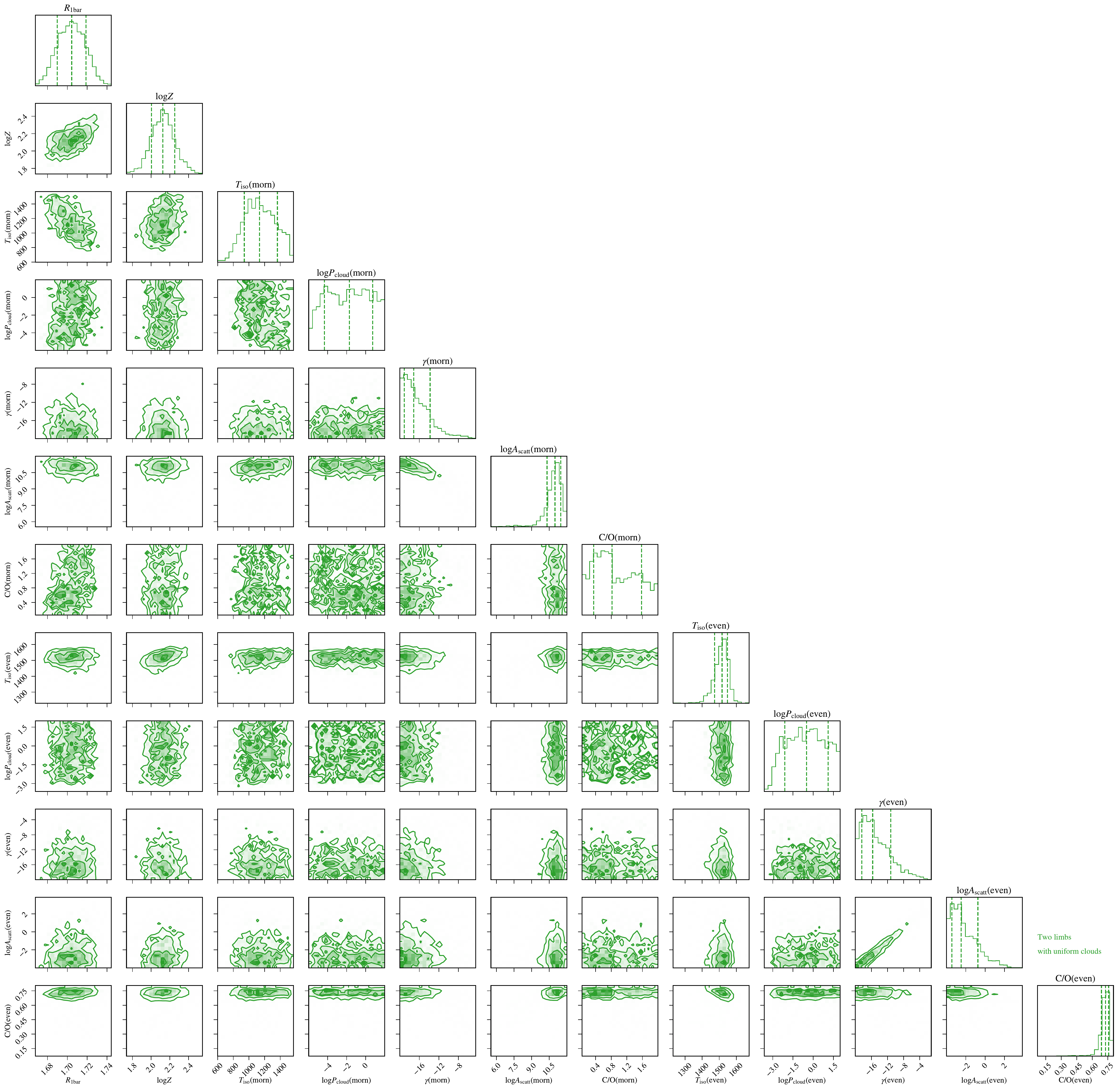}
\caption{Corner plots of the retrieval of Hypothesis 3, where two limbs with uniform clouds are assumed.}
\label{h3corner}
\end{figure*}

Figure \ref{h1corner}, \ref{h2corner}, and \ref{h3corner} show the joint posterior distributions of Hypotheses 1, 2, 3, respectively.

\begin{table}
\begin{center}
\caption[]{P200/COSMIC Transmission Spectrum of HAT-P-32Ab and the Adopted Dilution Flux Ratios}\label{tstable_p200}
 \begin{tabular}{ccc}
  \hline\noalign{\smallskip}
  \hline\noalign{\smallskip}
     Wavelength ({\AA}) &    $F_B/F_A$   &  $R_p/R_\star$   \\
  \hline\noalign{\smallskip}
    3990--4190 & 0.00051 & $0.1514 ^{+0.0030}_{-0.0034}$\\
    4190--4390 & 0.00047 & $0.1510 ^{+0.0024}_{-0.0025}$\\
    4390--4590 & 0.00066 & $0.1517 ^{+0.0042}_{-0.0039}$\\
    4590--4790 & 0.00063 & $0.1504 ^{+0.0022}_{-0.0020}$\\
    4790--4990 & 0.00079 & $0.1486 ^{+0.0018}_{-0.0018}$\\
    4990--5190 & 0.00077 & $0.1515 ^{+0.0013}_{-0.0013}$\\
    5190--5390 & 0.00137 & $0.1513 ^{+0.0019}_{-0.0018}$\\
    5390--5590 & 0.00150 & $0.1499 ^{+0.0016}_{-0.0016}$\\
    5590--5790 & 0.00169 & $0.1497 ^{+0.0016}_{-0.0016}$\\
    5790--5990 & 0.00148 & $0.1511 ^{+0.0015}_{-0.0015}$\\
    5990--6190 & 0.00232 & $0.1520 ^{+0.0016}_{-0.0017}$\\
    6190--6390 & 0.00183 & $0.1509 ^{+0.0015}_{-0.0016}$\\
    6390--6590 & 0.00311 & $0.1503 ^{+0.0013}_{-0.0013}$\\
    6590--6790 & 0.00263 & $0.1504 ^{+0.0017}_{-0.0018}$\\
    6790--6990 & 0.00281 & $0.1519 ^{+0.0015}_{-0.0016}$\\
    6990--7190 & 0.00379 & $0.1514 ^{+0.0017}_{-0.0018}$\\
    7190--7390 & 0.00556 & $0.1506 ^{+0.0017}_{-0.0018}$\\
    7390--7590 & 0.00846 & $0.1514 ^{+0.0018}_{-0.0018}$\\
    7590--7790 & 0.00673 & $0.1516 ^{+0.0025}_{-0.0028}$\\
    7790--7990 & 0.00821 & $0.1514 ^{+0.0030}_{-0.0030}$\\
    7990--8190 & 0.01062 & $0.1519 ^{+0.0025}_{-0.0027}$\\
    8190--8390 & 0.01113 & $0.1534 ^{+0.0022}_{-0.0023}$\\
    8390--8690 & 0.01090 & $0.1503 ^{+0.0030}_{-0.0031}$\\
    8690--8990 & 0.01286 & $0.1503 ^{+0.0026}_{-0.0025}$\\
    8990--9390 & 0.01462 & $0.1493 ^{+0.0022}_{-0.0022}$\\
  \noalign{\smallskip}\hline
\end{tabular}
\end{center}
\end{table}

\begin{table}
\begin{center}
\caption[]{Combined Optical Transmission Spectrum of HAT-P-32Ab}\label{tstable_combined}
 \begin{tabular}{cccc}
  \hline\noalign{\smallskip}
  \hline\noalign{\smallskip}
     Wavelength ({\AA}) & $R_p/R_\star$ \\
  \hline\noalign{\smallskip}
    3020--3300 & $0.15466 ^{+0.00189}_{-0.00189}$\\
    3300--3700 & $0.15287 ^{+0.00076}_{-0.00076}$\\
    3700--3950 & $0.15218 ^{+0.00061}_{-0.00061}$\\
    3950--4200 & $0.15141 ^{+0.00040}_{-0.00040}$\\
    4200--4350 & $0.15139 ^{+0.00056}_{-0.00056}$\\
    4350--4500 & $0.15146 ^{+0.00043}_{-0.00043}$\\
    4500--4650 & $0.15127 ^{+0.00054}_{-0.00054}$\\
    4650--4800 & $0.15122 ^{+0.00043}_{-0.00043}$\\
    4800--4950 & $0.15132 ^{+0.00041}_{-0.00041}$\\
    4950--5100 & $0.15115 ^{+0.00035}_{-0.00035}$\\
    5100--5250 & $0.15105 ^{+0.00035}_{-0.00035}$\\
    5250--5400 & $0.15138 ^{+0.00028}_{-0.00028}$\\
    5400--5550 & $0.15146 ^{+0.00027}_{-0.00027}$\\
    5550--5700 & $0.15164 ^{+0.00029}_{-0.00029}$\\
    5700--6000 & $0.15174 ^{+0.00018}_{-0.00018}$\\
    6000--6300 & $0.15130 ^{+0.00023}_{-0.00023}$\\
    6300--6500 & $0.15138 ^{+0.00037}_{-0.00037}$\\
    6500--6700 & $0.15129 ^{+0.00030}_{-0.00030}$\\
    6700--6900 & $0.15103 ^{+0.00031}_{-0.00031}$\\
    6900--7100 & $0.15113 ^{+0.00027}_{-0.00027}$\\
    7100--7300 & $0.15154 ^{+0.00026}_{-0.00026}$\\
    7300--7500 & $0.15121 ^{+0.00021}_{-0.00021}$\\
    7500--7700 & $0.15153 ^{+0.00027}_{-0.00027}$\\
    7700--8100 & $0.15129 ^{+0.00019}_{-0.00019}$\\
    8100--8350 & $0.15091 ^{+0.00024}_{-0.00024}$\\
    8350--8600 & $0.15063 ^{+0.00025}_{-0.00025}$\\
    8600--8850 & $0.15107 ^{+0.00030}_{-0.00030}$\\
    8850--9100 & $0.15070 ^{+0.00035}_{-0.00035}$\\
    9100--9500 & $0.15059 ^{+0.00038}_{-0.00038}$\\
    9500--10200 & $0.14903 ^{+0.00080}_{-0.00080}$\\
  \noalign{\smallskip}\hline
\end{tabular}
\end{center}
\end{table}

\end{document}